	\newfam\msyfam
	\font\tenmsy=msym10
	\font\eightmsy=msym8
 	\font\sixmsy=msym6
	\textfont\msyfam=\tenmsy
	\scriptfont\msyfam=\eightmsy
	\scriptscriptfont\msyfam=\sixmsy
	\def\matsymbol#1{{\fam\msyfam {#1}}}
	\newcommand{\be}{\begin{equation}}
        \newcommand{\ee}{\end{equation}}
        
        \newcommand{\ba}{\begin{eqnarray}}
        
        \newcommand{\lam}{\lambda}
        
        \newcommand{\tensor}{\otimes}
        \newcommand{\maps}{\colon}
        \newcommand{\ea}{\end{eqnarray}}
        \newcommand{\I}{{\bf I}}
        \newcommand{\K}{{\bf K}}
	\renewcommand{\L}{{\bf L}}
        \newcommand{\B}{{\cal B}}
	\newcommand{\T}{{\cal T}}
        \newcommand{\R}{{\matsymbol R}}
        \newcommand{\C}{{\matsymbol C}}

	\renewcommand{\H}{{\bf H}}

	\newcommand{\Diff}{{\rm Diff}}

	\documentstyle[12pt]{article}
\begin{document}

	\begin{center}
	{\bf Quantum Gravity and the Algebra of Tangles\\}
	\vspace{0.5cm}
	{\em John C. Baez\\}
	\vspace{0.3cm}
	{\small Department of Mathematics \\
	Wellesley College\\
	Wellesley, Massachusetts 02181\\
	(on leave from the University of California at Riverside)\\ }
	\vspace{0.3cm}
	{\small May 4, 1992}
	\vspace{0.5cm}
	\end{center}

\begin{abstract}  In Rovelli and Smolin's loop representation of
nonperturbative quantum gravity in 4 dimensions, there is a space of
solutions to the Hamiltonian constraint having as a basis isotopy
classes of links in $\R^3$.  The physically correct inner product on
this space of states is not yet known, or in other words, the
$\ast$-algebra structure of the algebra of observables has not been
determined.  In order to approach this problem, we consider a
larger space $\H$ of solutions of the Hamiltonian constraint,
which has as a basis isotopy classes of tangles.
A certain algebra $\T$, the ``tangle algebra,'' acts as operators on
$\H$.  The ``empty state''
$\psi_0$, corresponding to the class of the empty tangle, is
conjectured to be a cyclic vector for $\T$.    We construct simpler
representations of $\T$ as quotients of
$\H$ by the skein relations for the HOMFLY polynomial, and
calculate a $\ast$-algebra structure for $\T$ using these
representations.   We use this to determine the inner product of
certain states of quantum gravity associated to the Jones polynomial
(or more precisely, Kauffman bracket).
\end{abstract}

\section{Introduction}

In the Rovelli-Smolin approach to nonperturbative quantum
gravity in 4 dimensions \cite{RS,Smolin}, the ``kinematical
state space'' has as a basis generalized links in $\R^3$.
One obtains the physical state space by Dirac's procedure for
quantizing systems with constraints.  Thus, one first takes a quotient
of the kinematical state space by the
action of diffeomorphisms of $\R^3$ (more precisely, those connected
to the identity).  Then, inside this quotient space
one seeks solutions to the Hamiltonian constraint; these form the
physical state space.  A large set of solutions is known,
corresponding simply to isotopy classes of unoriented links in $\R^3$.
While there is much to be done towards finding the
complete solution space of the Hamiltonian constraint, and clarifying
technical issues related to regularization and operator ordering,
 the main ``practical'' problem is that the inner product on the
physical state space
is not known; it is presently merely a vector space, not a Hilbert space.
(Indeed, it is not even known whether isotopy classes of unoriented links
are truly normalizable elements of the physical Hilbert space,
or whether one must form ``wave packets'' to obtain vectors of finite
norm.)

This is important because an inner product is essential in the
probabilistic interpretation of a quantum theory.  Alternatively, one
could say that the problem is the lack of an adjoint on the algebra of
linear operators on the physical state space.  Recall that an
algebra $A$ is a {\it $\ast$-algebra} if there is a map $\ast \maps A
\to A$ such that
\[    (a^\ast)^\ast = a,\quad (\lambda a)^\ast = \overline \lambda a^\ast,
\quad (a + b)^\ast = a^\ast + b^\ast, \quad (ab)^\ast = b^\ast a^\ast .\]
 Until an operator algebra
has been given the structure of a $\ast$-algebra, one can identify
neither the observables (self-adjoint elements) nor
the physical states
(functionals $\psi \maps A \to \C$ such that $\psi(1) = 1$ and
$\psi(a^\ast a) \ge 0$ for all $a \in A$).   Thus the algebra of
operators for a physical system should be a $\ast$-algebra.

For reasons of mathematical convenience and physical principle,
it is usually assumed that the algebra of
observables is a C*-algebra.
Recently, beginning with the work of Jones \cite{Jones},
a profound connection has been found between C*-algebras, especially
type ${\rm II}_1$ factors, and link invariants.  In this paper we
attempt
to exploit this connection to shed some light on the $\ast$-algebraic
aspects of quantum gravity.

We begin by considering the structure of quantum gravity at spacelike
infinity.  Our treatment will be brief, as it is a natural
generalization of the original work by Rovelli and Smolin,
and it will only serve as a heuristic preparation
for the material in the next section.   We fix
an asymptotically flat structure on $\R^3$, and
instead of working only with links
as states, we consider a generalization of links in which certain
strands extend to infinity in an asymptotically geodesic fashion
in $\R^3$.   These will be regarded as embedded in the
closed unit ball, $D^3$, and are a slight variation on what are commonly
known in knot theory as tangles.  In an obvious generalization
of their original construction, there is a representation of the
Rovelli-Smolin quantized loop observables on the vector space
built up from generalized tangles having certain types of
self-intersection.  This is the {\it kinematical} state space.
The key point is that the loop observables are ``local,'' that is,
they do not affect the structure of the tangles at
spacelike infinity.  Thus the geometry of a tangle at spatial
infinity, which in the next section we call ``boundary data,''
defines superselection sectors of the kinematical state space.

Then, rather than taking the quotient by all
diffeomorphisms of $\R^3$, we do so only by those
that extend to diffeomorphisms of $D^3$ equal
to the identity on the boundary, $\partial D^3 = S^2$.
This yields the {\it diffeomorphism-invariant} state
space, which has as its basis isotopy classes of tangles with
intersections, where isotopies are required to leave $S^2$ fixed.
Finally, vectors in the diffeomorphism-invariant state space
that are annihilated by the Hamiltonian constraint span the {\it
physical} state space.  More precisely, we only require that
\[       (\int_{\R^3} H(x) f(x) d^3 x) \psi = 0 \]
for compactly supported $f$, so that the analysis of Rovelli and
Smolin extends to the case of framed tangles.  In particular, isotopy
classes of honest framed tangles (with no intersections) span a
subspace of the space of physical states.  For the most part
we only consider such states.    Thus we work with a theory whose
state space $\widetilde \H$ has as a basis isotopy classes of tangles.

While somewhat technical, it is important here to mention the issue of
equipping tangles with framings and orientations.
Rovelli and Smolin's original construction worked with unoriented links in
$\R^3$.   More recent work of Br\"ugmann, Gambini and Pullin \cite{BGP1}
 suggests that framed links are required to deal with
regularization issues.  Their work is closely related to the need for framings
in Chern-Simons theory \cite{Witten}.  In work whose relation to the
above is not yet quite clear, Ashtekhar and Isham \cite{AI} have
introduced framed links, or more precisely ``strips,'' in a rigorous
treatment of some aspects of the loop representation.  Certainly, to
make contact with the Reshetikhin-Turaev theory of link invariants
\cite{RT,Turaev}, framings should be taken into account.
In addition to framings,
 orientations may be required in theories of gravity
coupled to matter.   We thus take as a basis for $\widetilde \H$
isotopy classes of framed oriented tangles.   This is probably the
most economical approach in the long run, as the modifications necessary to
treat the unframed or unoriented cases are easy, and tangles of all
types are a promising approach to a central problem of quantum gravity:
describing local excitations in a manifestly diffeomorphism-invariant
manner.

In Section 2 we develop a framework for handling symmetries and
other natural operators on the space $\widetilde \H$.
The group of orientation-preserving diffeomorphisms of $S^2$ acts
on $\widetilde \H$, and one may mod out by
almost all of this group to obtain a reduced state space $\H$.
The remaining symmetries are described by a discrete group, the
``tangle group.''  The action of this group together with certain
annihilation and creation operators generates an algebra we call the
``tangle algebra.''
It seems that the whole space $\H$ may be
built up from a certain vector $\psi_0$, somewhat analogous to the vacuum in
ordinary quantum field theory, by applying operators in the tangle
algebra.

The next step is to determine the physically correct
 $\ast$-structure of the tangle algebra.
Since the elements of the tangle group represent symmetries, it is
natural to assume that they are unitary.  The difficulty is to find
the adjoints of the annihilation and creation operators.  To gain
insight into this issue, in Section 3 we consider simpler
representations of the tangle algebra, or ``tangle field theories,''
 obtained as quotients of $\H$ by certain relations, essentially the
skein relations for the HOMFLY polynomial.  The HOMFLY polynomial is
closely related to the representation theory of the quantum groups
$SL_q(n)$, and for unitarity of the tangle group action $q$ must be a
root of unity.  In fact, unitarity imposes enough constraints on these
tangle field theories to calculate the adjoint of the
annihilation operator in terms of the creation operator.  As a special
case of these tangle field theories, we obtain an unoriented tangle
field theory corresponding to the Kauffman bracket (a
normalized form of the Jones polynomial).  This theory
may be regarded as a reduction of quantum gravity, and states of this
theory correspond to the states of quantum gravity obtained via
Chern-Simons theory by Br\"ugmann, Gambini and Pullin \cite{BGP1}.
Our calculations essentially determine the inner product of such
states from the unitarity of the tangle group action.
In Section 4 we sketch a general method of obtaining tangle field theories
from representations of quantum groups.  A
systematic investigation of these tangle field theories should shed
more light on the $\ast$-algebraic aspects of quantum gravity.

The author wishes to acknowledge Dan Asimov, Dror Bar-Natan,
Greg Kuperberg, Curtis McMullen,
Geoffrey Mess, Jorge Pullin, J\'ozef Przytycki,
and Stephen Sawin for helpful discussions.

\section{The Tangle Algebra}

We define a {\it tangle} in a manifold with boundary $M$
to be an oriented smooth
$1$-dimensional submanifold $X$ of $M$, possibly with boundary,  such that
$\partial X\subset \partial M$ and such that $X$ meets $\partial M$
transversally.
A tangle $X$ is thus a disjoint union of connected components, which
are oriented circles contained in the interior of $M$ or oriented
paths connecting two points on $\partial M$.
The connected components of a tangle will be called {\it strands}, and
call the points in $\partial X$ {\it boundary points}.
Using the orientation on $X$, we
may describe these boundary points as either {\it incoming} or {\it
outgoing} points.   If there are $n$ incoming boundary
points, there are $n$ outgoing boundary points, and we call $n$ the {\it
boundary number} of the tangle.

Now suppose that $M$ is a 3-manifold.
A {\it framing} of a tangle $X \subset M$ is a smooth section $v$ of the
tangent bundle $TM$ over $X$ such that for all $x \in X$, $v_x \notin
T_xX$; for boundary points $x \in X$ we require that $v_x$ is tangent
to $\partial M$.  A framed tangle may be visualized as a disjoint union of
ribbons.  Diffeomorphisms of $M$ act on framed tangles in $M$ in a
natural way.
 We say two framed tangles $X$ and $X'$ are {\it isotopic} if there is
a continuous one-parameter family $f_t$ of diffeomorphisms of $M$ such
that $f_0$ is the identity, $X' = f_1(X)$, and
$f_t$ is the identity on $\partial M$ for all $t$.

Usually knot theory considers tangles in $[0,1] \times \R^2$.
For us, tangles will be taken in the ball $D^3$ unless otherwise specified.
Let $\widetilde \H$ denote the vector space having as its basis isotopy classes
of framed tangles.  Note that
\[   \widetilde \H = \bigoplus_{n=0}^\infty \widetilde \H_n   \]
where $\widetilde \H_n$ is spanned by isotopy classes
of framed tangles with boundary number $n$.    We may further
decompose the spaces $\widetilde \H_n$ as follows.
For any tangle with boundary number $n$,
we write the incoming points as $x^- = (x_1^-, \dots, x_n^-)$ and the outgoing
points as $x^+ = (x_1^+, \dots, x_n^+)$.  We have $x = (x^-,x^+) \in
(S^2)^{2n} - \Delta$, where $\Delta$ is the set of $2n$-tuples of
points in $S^2$ at least two of which are equal.
To each framed tangle we may associate the pair
$(x,v)$, where $x \in (S^2)^{2n} - \Delta$ is as above, and
\[  v = (v_{x_1^-}, \dots, v_{x_n^-}, v_{x_1^+}, \dots v_{x_n^+}) .\]
However, this association is not canonical, since we may choose any
ordering of the incoming/outgoing boundary points.
Let ${\cal C}(n)$ denote the space of
pairs $(x,v)$ such that $x \in (S^2)^{2n} - \Delta$
and $v = (v_1^-, \dots, v_n^-, v_1^+, \dots, v_n^+)$, where $v_i^\pm$ is
a nonzero vector in $T_{x_i^\pm}S^2$.   Let $S_n$ denote the symmetric group.
A pair $(\sigma, \tau) \in S_n \times S_n$ acts on
$(x,v) \in {\cal C}(n)$ by permuting the incoming boundary points
$x_i^-$ and tangent vectors $v_i^-$ with $\sigma$, and the outgoing
boundary points $x_i^+$ and tangent vectors $v_i^+$ with $\tau$.
Thus to each framed tangle we may canonically associate {\it boundary data}
$[x,v] \in \B(n)$, where $\B(n)$ is the quotient space ${\cal C}(n)/(S_n
\times S_n)$.  Each space $\widetilde \H_n$ is thus a direct sum
\[   \widetilde \H_n = \bigoplus_{[x,v] \in \B(n)}   \H_n[x,v]   ,\]
where $\H_n[x,v]$ denotes the space spanned by isotopy classes of
framed tangles with boundary data $[x,v] \in \B(n)$.
The reader may find this uncountable direct sum surprising.
It is, of course, required by the fact that $\H$ has an uncountable
basis, while each $\H_n[x,v]$ has a countable basis.  It seems likely
that in a more analytically sophisticated approach this direct sum would be
replaced by a direct integral.   Here, however, we take advantage of
the fact that the spaces $\H_n[x,v]$ are the fibers of a flat vector bundle
over $\B(n)$.

To see this, first note that each diffeomorphism of $S^2$ extends to a
diffeomorphism of $D^3$, unique up to isotopy.
This intuitive but highly nontrivial result is due to Cerf, Munkres,
and Smale \cite{Cerf,Munkres,Smale}.  It follows that
$\Diff^+(S^2)$, the group of orientation-preserving
diffeomorphisms of $S^2$, acts on isotopy classes of framed
tangles.  There is thus a representation $\rho$ of $\Diff^+(S^2)$ on
$\H$.  Note that if $g \in \Diff^+(S^2)$,
\[           \rho(g) \maps \H_n[x,v] \to \H_n(g[x,v]) \]
where we define
\[    g[x,v] = [g(x_1^-), \dots, g(x_n^+), dg(v_1^-), \dots, dg(v_n^+)] .\]
It follows that for any $[x,v], [x',v'] \in \B(n)$, we may identify
the spaces $\H_n[x,v]$ and $\H_n[x',v']$.  This
identification is not unique, since there are many $g \in \Diff^+(S^2)$
with $g[x,v] = [x',v']$.   However, if $[x,v]$ and $[x',v']$ are close,
we may take any $g \in \Diff^+(S^2)$ sufficiently close to the
identity with $g[x,v] = [x',v']$,
and identify $\H_n[x,v]$ with $\H_n[x',v']$ via $\rho(g)$.
The map $\rho(g)$ does not depend on the choice of such $g$, so the
spaces $\H_n[x,v]$ are the fibers of a flat vector bundle over
$\B(n)$.

This device reduces the study of the
state space $\widetilde \H$ to the study of the spaces $\H_n[x,v]$,
where for each $n \ge 0$, $[x,v]$ is a single arbitrary element of $\B(n)$.
Thus we fix once and for all distinct points $x_i^\pm$ on $S^2$
and nonzero tangent vectors $v_i^\pm \in T_{x_i^\pm}S^2$ for
$i = 1,2,3,\dots$, and let
\[       \H_n = \H_n[x,v] \]
where $x = (x_1^- , \dots, x_n^-, x_1^+, \dots, x_n^+)$ and
$v = (v_1^- , \dots, v_n^-, v_1^+, \dots, v_n^+)$.
We may picture the points $x_1^-, x_2^-, \dots$ as
lined up from left to right near the north pole of $S^2$, and
$x_1^+, x_2^+, \dots$ as lined up from left to right near the south
pole, with the tangent vectors $v_i^\pm$ pointing to the right.
Of course, the actual geometry of the situation is irrelevant, as any
choice of points may be brought into this position by a diffeomorphism.
We let
\[      \H = \bigoplus_{n=0}^\infty \H_n  .\]

By forming the space $\H_n$, we have almost reduced the space
$\widetilde \H_n$ by the action of all orientation-preserving
diffeomorphisms of $D^3$.   However, diffeomorphisms which fix the
equivalence class $[x,v]$ still act as
symmetries of $ \H_n$.   Indeed, even a diffeomorphism which fixes
$(x,v)$ can act nontrivially on $\H_n$.  Since the space
$\H_n[x,v]$ is the fiber of a flat vector bundle over $\B(n)$,  the
{\it tangle group}
\[       T_n = \pi_1(\B(n))  \]
acts on $\H_n$ by holonomy.  This discrete symmetry group action
on $\H_n$ captures the diffeomorphism-invariance of the space
$\widetilde \H$.   We denote the representation of $T_n$ on $\H_n$ by
$\rho_n$, and we will also regard the operators $\rho_n(g)$ as
operators on $\H$ by extending them to be zero on all
$\H_m$ with $m \ne n$.

An element of $T_n$ may be represented as a loop in
the space $\B(n)$, or as a path in ${\cal C}(n)$.  It may thus be
represented as a framed tangle
in $[0,1] \times S^2$ with boundary number $2n$, having the points
$(1,x_i^-)$ and $(0,x_i^+)$ as incoming and $(0,x_i^-)$ and
$(1,x_i^+)$ as outgoing boundary points.
Each strand must go either from a point of the form $(1,x_i^-)$ to
one of the form $(0,x_j^-)$, or from one of the form $(0,x_i^+)$ to one
of the form $(1,x_j^+)$.
The framing at the boundary points must match up with the
standard vectors $v_i^\pm$ in the obvious way.  Moreover, the tangle
cannot contain any embedded circles, and all the embedded line
segments must be, up to orientation, of the form
\[          t \mapsto (t,f(t)) \]
for some $f \maps [0,1] \to S^2$.  This sort of framed tangle is, in fact,
a particular sort of framed braid on $S^2$.
The action of an element of $T_n$ on
an element of $\H_n$ is illustrated in Figure 1.  In the case of a
theory of unoriented framed tangles we would enlarge the symmetry
group from $T_n$ to the whole framed braid group $FB_{2n}(S^2)$, since
there would no longer be a fundamental distinction between incoming
and outgoing strands.  In the case of a theory of tangles with neither
orientation nor framing the symmetry group would just be the group
$B_{2n}(S^2)$ of braids on $S^2$.  (See \cite{Baez} for the definition
of these groups.)

Let the {\it empty state}, $\psi_0 \in \H_0$, be the basis vector
corresponding to empty set, which is vacuously a
$1$-dimensional submanifold of $D^3$.
The state $\psi_0$ is not
a nondegenerate ground state of a Hamiltonian, since all states in
$\H$ are annihilated by the Hamiltonian constraint.  However,
$\psi_0$ is analogous to the vacuum,
in that many, perhaps all,
vectors in $\H$ may be obtained from $\psi_0$
by applying certain operators and taking linear combinations, as we
now describe.

The {\it creation} operator
\[    c \maps \H_n \to \H_{n+1}   \]
simply adds an extra strand to any tangle with boundary number $n$;
the new strand is required to be unknotted, untwisted,
 and to remain to the right
of the existing strands.  Figure 2 shows a tangle representing
 $\psi \in \H_n$ and the tangle representing $c\psi$.   For
$n \ge 1$, the {\it annihilation} operator
\[   a \maps \H_n \to \H_{n-1}  \]
moves the rightmost boundary points, $x_n^-$ and $x_n^+$, slightly
into the interior of $D^3$, and connects them with a smooth arc that
is unknotted, untwisted,
and remains to the right of the existing strands.   We
define $a$ to be zero on $\H_0$.
Figure 3 shows a tangle representing
$\psi \in \H_n$ and the tangle representing $a \psi$.
It is essential, but easy, to check that the annihilation and creation
operators are well-defined.

We call the algebra of operators on $ \H$ generated by the
annihilation and creation operators together with the operators
representing the tangle groups the {\it tangle algebra}, $\T$.
It is a consequence of Alexander's theorem \cite{Birman}
that all isotopy classes of framed oriented
links may be written as $a^n \rho_n(g)c^n  \psi_0$ for some
sufficiently large $n$, where $g \in T_n$.
It follows that every vector in $\H_0$ is of the form $A\psi_0$
for some $A \in \T$.   We conjecture that in fact
 every vector in $\H$ is of this form, i.e., that the empty state
is a cyclic vector for the tangle algebra.

Simple physical
considerations give some information about the Hilbert space structure
of $\widetilde \H$ and the reduced space $\H$.
First, it is natural to assume that the representation
$\rho$ of $\Diff^+(S^2)$ on $\widetilde \H$
is unitary, as it acts as symmetries.  Since the spaces $\widetilde \H_n$ are
superselection sectors it is also natural to assume that they are
 orthogonal.  At the level of reduced state spaces, we may assume that
each representation $\rho_n$ is unitary and the direct sum decomposition
$\H = \bigoplus \H_n $ is orthogonal.
Let $p_n \maps \H \to \H_n$ be the orthogonal projection, so that
\[   p_n = p_n^\ast, \qquad  p_np_m = \delta_{nm} p_n .\]
Then for all $g \in T_n$ we have
\[       \rho_n(g)^\ast = \rho_n(g^{-1})   ,\]
\[     \rho_n(g)\rho_n(g)^\ast =
 \rho_n(g)^\ast \rho_n(g) = p_n .\]
It follows that to determine a $\ast$-algebra structure of $\T$, it
suffices to describe the adjoints $a^\ast$ and $c^\ast$.  Of course, it is
possible that these do not lie in $\T$, in which case $\T$ must be
extended by the operators $a^\ast$, $c^\ast$ to become a $\ast$-algebra.
In the next section we present a simplified model in which
$a^\ast$ is a constant times $c$.
This, of course, further justifies the analogy with
annihilation and creation operators.

\section{Tangle Field Theories}

It is plausible that in some sense, quantum gravity should be
4-dimensional topological quantum field theory.  A
definition of topological quantum field theories has been given by
Atiyah \cite{Atiyah}, and examples have been constructed in 3
dimensions \cite{RT2,TV}.   However, one expects quantum gravity to differ
considerably from
3-dimensional topological quantum field theories, since in
gravity there are local excitations.  This is reflected in the fact
that the reduced state space $\H$ is infinite-dimensional, with
isotopy classes of tangles representing local excitations.
Additionally, we have seen that $ \H$ is a {\it tangle field
theory}, that is, a representation of the tangle algebra $\T$.
It seems that some insight into quantum gravity may be gained by
considering simpler tangle field theories formed as quotients of
$\H$.  As an example, we construct a
family of tangle field theories related to the HOMFLY
polynomial invariant of links.   As shown by Reshetikhin and Turaev
\cite{RT,Turaev}, this invariant is associated to the quantum groups
$SL_q(n)$.  We obtain tangle field theories that are unitary
representations of the tangle group only when the parameter
$q$ is a root of unity.

Let $t,x,y$ be arbitrary nonzero complex numbers.
For $n \ge 1$, let $\I_n$ be the subspace of $ \H_n$ spanned by the elements
\be    t^{-1}y^{-1}\phi_+ - ty\phi_- - x\phi_0 ,
\label{skein}\nonumber\ee
where $\phi_+, \phi_-, \phi_0 \in \H_n$ are the
isotopy classes of three framed tangles with identical pictures except
within a small disk, where they appear as in Figure 4, and
\be    y\psi - \psi' ,
\label{twist}\nonumber\ee
where $\psi, \psi' \in \H_n$ are identical except
within a small disk, where they appear as in Figure 5.
Let $\I_0$ be the subspace of $\H_0$ be the space spanned by the
elements (\ref{skein}) and (\ref{twist}), together with
\be \psi_{unknot} - {t^{-1} - t\over x} \psi_0 ,  \label{unknot}  \ee
where $\psi_{unknot} = ac\psi_0 \in \H_0$ is the
isotopy class of an unknotted circle in $D^3$.
Let  $\K_n =  \H_n/\I_n$.   It is easy to
check that the representation of $\T$ on $\H$ gives rise to a
representation of $\T$ on the space
\[  \K = \bigoplus_{n=0}^\infty \K_n  .\]
Thus $\K$ is a tangle field theory formed as quotient of $\H$.
For any $\psi \in \H$, we write $[\psi]$ for the corresponding
vector in $\K$.

It is easy to calculate in the tangle field theory $\K$, because the
relations (\ref{skein}) and (\ref{twist}) are the skein relations for the
HOMFLY polynomial invariant of links \cite{HOMFLY}
(normalized as in the paper by Jones \cite{Jones}) multiplied by
 $y^w$, where $w$ is the writhe of
the link in question.   For any nonempty framed link $\psi \in \H_0$,
repeated use of these relations, together with (\ref{unknot}),
allows one to express $[\psi] \in \K_0$ as a
multiple of $[\psi_0]$.

Next, note that each space $\H_n$ is an algebra, with
the product $\psi \psi'$ of two framed
tangles $\psi$ and $\psi'$ given by attaching
the $i$th outgoing boundary point of $\psi'$ to the $i$th incoming
boundary point of $\psi$.  (This remarkable fact, that a space of
states should have an algebra structure, also holds for the state
space of $S^n$ in any topological quantum field theory satisfying
Atiyah's axioms.)  The identity of $\H_n$ is the vector
\[      \psi_n = c^n \psi_0 .  \]
Moreover, the subspace $\I_n$ is an ideal of $\H_n$, so
the quotient $\K_n$ inherits an algebra structure
from $\H_n$.   We can use this to calculate $\K_n$ for $n \ge 0$, as follows.

The tangle group $T_n$ has the braid
group $B_n$ as a subgroup.  Recall that $B_n$ is generated by elements
$s_i$, $1 \le i < n$, with relations
\ba    s_is_j &=& s_j s_i \qquad\qquad\qquad |i-j| > 1, \nonumber\cr
s_is_{i+1}s_i &=& s_{i+1}s_i s_{i+1} .\nonumber\ea
We may regard $B_n$ as a subgroup of $T_n$ by letting $s_i$ switch
the $i$th and $(i+1)$st outgoing strands as in Figure 6.
The braid group action on $\H_n$ is related to its algebra structure
as follows:
\[ \rho_n(gh)\psi_n = (\rho_n(g)\psi_n)(\rho_n(h)\psi_n)    \]
 for any $g,h \in B_n$.
We may use the skein relations (\ref{skein}) to show that
elements of the form $[\rho_n(g)\psi_n]$ span $\K_n$, where $g \in B_n$.
Moreover, the skein relations imply that
\[   [\rho_n(s_i^2)\psi_n] = tyx [\rho_n(s_i)\psi_n] + (ty)^2[\psi_n]   .\]

Let $H(q,n)$ denote the Hecke
algebra with generators $g_i$, $1 \le i < n$, and relations
\ba    g_ig_j &=& g_j g_i \qquad\qquad\qquad |i-j| > 1, \nonumber\cr
g_ig_{i+1}g_i &=& g_{i+1}g_i g_{i+1}, \nonumber\cr
g_i^2 &=& (q-1)g_i + q. \nonumber\ea
We define $H(q,1)$ to be $\C$.
Suppose that
\[       ty = \lam^{1/2} q^{1/2}  ,\qquad
x = q^{1/2} - q^{-1/2}  ,\]
where we choose, once and for all, square roots of $\lam$ and $q$.
Then for there is clearly a homomorphism
\[                H(q,n) \mapsto \K_{n-1}  \]
given by
\[                g_i \mapsto \lam^{-1/2} [\rho_n(s_i)\psi_n]  .\]
In fact, it follows from results of Turaev \cite{Turaev2} that this is an
isomorphism.  As we shall show, in certain cases
this isomorphism gives rise to a natural inner product on a quotient
\[  \L = \bigoplus_{n=0}^\infty \L_n  \]
of $\K$.  The summands $\L_n$ are orthogonal with
respect to this inner product, and the tangle group $T_n$ acts as
unitary operators on $\L_n$.  These are precisely the properties
we argued for on physical grounds in the previous section, and should
probably be added to the definition of a tangle field theory.
Moreover, relative to this inner product, $a^\ast$ is just a scalar
multiple of $c$.

As shown by Ocneanu \cite{HOMFLY} and Wenzl \cite{Wenzl},
for any $z \in \C$ there exists a unique trace $\tau$ on the inductive limit
of the Hecke algebras such that $\tau(1) =1$ and the Markov property
 $\tau(g_nx) = z\tau(x)$ holds for all $x \in H(q,n)$.
Using the isomorphism above we obtain traces $\tau_n$ on the algebras
 $\K_n$ determined by the properties
\ba          \tau_n([\psi_n]) &=& 1   \nonumber\cr
         \tau_{n+1}([c\psi]) &=& \tau_n([\psi])  \nonumber\cr
         \tau_{n+1}([\rho_{n+1}(s_n)c\psi]) &=&  \lambda^{1/2}z\tau_n([\psi])
\nonumber\ea
for all $n$ and $\psi \in \H_n$.  Note that
\[\tau_{n+1}([\rho_{n+1}(s_n^{-1})c\psi]) =  \lambda^{-1/2}(q^{-1}z
+ q^{-1} - 1)\tau_n([\psi]) .\]
If
\be        y = (\lambda z/(q^{-1}z + q^{-1} - 1))^{1/2}  \label{y} \ee
and we define
\be     N =  (z(q^{-1}z + q^{-1} - 1))^{1/2}  \label{N} \ee
with a consistent choice of square roots, we thus have
\ba \tau_{n+1}([\rho_{n+1}(s_n)c\psi]) &=&  yN \tau_n([\psi]) ,\nonumber\cr
\tau_{n+1}([\rho_{n+1}(s_n^{-1})c\psi]) &=&  y^{-1}N \tau_n([\psi]) .
\nonumber\ea
These equations describe how the Markov moves affect the trace.

Henceforth we assume that equation (\ref{y}) holds.
Note that it implies $N^{-1} = (t^{-1} - t)/x$, so that
\be        \psi_{unknot} = N^{-1} \psi_0   . \label{unknot2}  \ee
 By the theory relating link and tangle invariants to Markov traces
\cite{Jones,Turaev,Turaev2}, it follows that
\[     {[a^n\psi]\over [\psi_0]} =  N^{-n}  \tau_n([\psi])     \]
for all $\psi \in \H_n$.  Moreover, if $\psi$ is the isotopy class of
a framed tangle, the quantity $[a^n\psi]/[\psi_0]$
depends only on the isotopy class of the framed link $a^n\psi$, the
closure of $\psi$.

When $|q| = 1$ there is a unique $\ast$-structure on $H(q,n)$
such that $g_i^\ast = g_i^{-1}$.  We transfer this to $\K_n$, making
it into a $\ast$-algebra.   Note that if $ |\lam| = 1$,
\[       [\rho_n(g)\psi_n]^\ast = [\rho_n(g^{-1})\psi_n]  .\]
If in addition $|y| = 1$, the relation (\ref{twist})
implies that for $\psi\in \H_n$ the isotopy class of a tangle,
\[           [\psi]^\ast = [\psi^\ast]  ,\]
where $\psi^\ast\in \H_n$ is the isotopy class of the tangle with
the opposite orientation, reflected about the $xy$-plane
 as in Figure 7.    Here we assume, without loss of generality,
that reflection of the points
$x_i^+$ and vectors $v_i^+$ about the $xy$-plane yields
$x_i^-$ and $v_i^-$, respectively.  This operation on
tangles extends uniquely to a $\ast$-structure on $\H_n$.

As shown by Ocneanu and Wenzl, the traces $\tau_n$ are positive, that is,
\[     \tau_n(\psi^\ast \psi) \ge 0  \]
for all $\psi \in  \K_n$, if and only if $q$ and $z$ satisfy
\[ q = e^{\pm 2\pi i/\ell}  ,\qquad  z = {q-1\over 1 - q^k},  \]
where $k,\ell$ are integers with $0 < k < \ell$.
{}From now on we assume these conditions on $q$ and $z$, and assume
$|\lambda| = 1$.  It follows
from equations (\ref{y}) and (\ref{N}) that
\[   N = {q^{-k/2} - q^{k/2}\over q^{1/2} - q^{-1/2}}
 , \qquad  y = \lambda^{1/2} q^{(1-k)/2}  .\]

We give the space $\K_n$ an ``inner product'' by setting
\[  \langle \psi,\phi \rangle
  = \left\{ \begin{array}{ll}
 \tau_n(\psi^\ast \phi)         &{\rm if}\; n = m        \\
 0                              &{\rm if}\; n \ne m
\end{array} \right.         \]
for $\psi\in \K_n$ and $\phi\in\K_m$.
This has all the properties of a true inner product except that it is not
definite, that is, there are typically nonzero states $\psi \in \K_n$
with $\langle \psi,\psi \rangle = 0$.  We deal with these states of
norm zero later; first we show that the tangle group action preserves
this ``inner product'' on $\K_n$ and compute $a^\ast$.

Let $\psi, \phi \in \H_n$ be isotopy classes of tangles.
To show that the tangle group $T_n$ preserves the ``inner product'' on $\K_n$
it suffices to note that
\[            \tau_n([(\rho_n(g)\psi)^\ast (\rho_n(g)\phi)]) =
\tau_n([\psi^\ast\phi])  \]
for any $g \in T_n$.  In fact, the isotopy classes
$(\rho_n(g)\psi)^\ast (\rho_n(g)\phi)$ and $\psi^\ast \phi$ are equal.
This is easily seen using pictures.
We also claim that
\[   \langle \psi, c\phi\rangle =  N\langle a\psi,\phi\rangle \]
for all $\psi \in \K_n$ and $\phi \in \K_{n-1}$.
It suffices to consider the case where
\[     \psi = [\rho_n(g)\psi_n],\qquad \phi =
[\rho_{n-1}(h)\psi_{n-1}], \]
for $g \in B_n$ and $h \in B_{n-1}$.  We have
\[     \langle \psi, c\phi\rangle =
 N^n [a^n((\rho_n(g)\psi_n)^\ast (c\rho_{n-1}(h)\psi_{n-1}))]/[\psi_0]. \]
{}From Figure 8, it is easy to see that
\[    a^n((\rho_n(g)\psi_n)^\ast (c\rho_{n-1}(h)\psi_{n-1})) =
a^{n-1}((a\rho_n(g)\psi_n)^\ast (\rho_{n-1}(h)\psi_{n-1})) .\]
It follows that
\[  \langle \psi, c\phi\rangle = N^n [a^{n-1} (a(\rho_n(g)\psi_n)^\ast
(\rho_{n-1}(h)\psi_{n-1}))]/[\psi_0] = N \langle a\psi,\phi \rangle  .\]
We thus have $a^\ast = N^{-1}c$.

Finally, to obtain a tangle field theory with a positive definite
inner product, we simply define $\L_n$ to be the quotient of $\K_n$ by
the subspace of $\psi \in \K_n$ such that $\langle \psi,\phi \rangle =
0$ for all $\phi \in \K_n$.  The space $\L_n$ inherits an inner
product from $\K_n$, and
since $\L_n$ is finite-dimensional it is a Hilbert space.
Let $\L$ denote the Hilbert space direct sum of the spaces $\L_n$.
Since the tangle group $T_n$ preserves the ``inner product'' on $\K_n$,
it has a
unitary representation on $\L_n$.  Note also that if $\psi \in \K$
has the property that $\langle \psi,\phi \rangle = 0$ for all $\phi
\in \K$, the vectors $a\psi$ and $c\psi$ share this property, since
\[          \langle \psi, c\phi\rangle = N \langle a\psi,\phi\rangle .\]
It follows that the operators $a,c$ on $\K$ define operators on the
quotient space $\L$, and $\L$ becomes a representation of the whole
tangle algebra.  In addition to the relation $a^\ast = N^{-1}c$, it is
worth noting that the operator $ac$, which adds an extra unknotted
circle to any tangle, satisfies
\[         ac = N^{-1} .\]
To show this holds when applied to $\psi_0$ requires equation
(\ref{unknot2}).

If we choose $k = 2$ and $\lambda^{1/2} = -q^{1/4}$, the link invariant
$[a^n\psi]/[\psi_{unknot}]$ is equal to the
the Kauffman bracket \cite{Kauffman}, which is just the Jones
polynomial times $(-q)^{-3w/4}$, where $w$ is the writhe.   The Kauffman
bracket is an
invariant of unoriented framed links.  Thus in this case we obtain an
unoriented tangle field theory, hence a reduction of quantum gravity.
Moreover, in this case $\L_n$ may be identified with the
Temperley-Lieb algebra \cite{Jones}.
It should be noted that the Kauffman bracket
is implicit in the work of Br\"ugmann, Gambini and Pullin \cite{BGP1},
who obtain the Jones polynomial times a function of
the writhe when constructing states of quantum gravity from
Chern-Simons theory.  The above results effectively determine the
inner product on ``Chern-Simons states'' of quantum gravity on $D^3$ from the
unitarity of the tangle group action.

\section{Conclusions}

While the above argument obtains an inner product on $\L$,
hence a $\ast$-structure for
the tangle algebra as represented on $\L$,
it would be preferable to have physical grounds for choosing an inner product
on $\H$, or its unoriented analog.  The problem, of course, is that we have
little idea what the physical observables are in a manifestly
diffeomorphism-invariant formulation of quantum gravity.  Thus it seems
worthwhile to examine a variety of other tangle field theories formed as
reductions of $\H$.  For example, there should be an unoriented tangle field
theory based on skein relations for the Kauffman polynomial, in which the
Birman-Wenzl algebra takes the place of the Temperley-Lieb algebra \cite{BW}.

More generally,
given a representation $V$ of a quantum group (or more precisely,
ribbon Hopf algebra), the work of Reshetikhin
and Turaev \cite{RT,Turaev,Turaev2} shows how to associate to any
element of $\H_n$ a linear transformation of $V^{\tensor n}$.
If we let $\K_n$ denote the range of $\H_n$ in
the space of linear transformations ${\rm Hom}(V^{\tensor n})$, then
$\K = \bigoplus \K_n$ becomes a
tangle field theory by the methods of the previous section.
In particular, the HOMFLY tangle
field theories are associated to the quantum groups $SL_q(n)$, while
the Kauffman tangle field theories are associated to $SO_q(n)$ and $Sp_q(n)$.
 To further understand the
$\ast$-algebraic aspects of the tangle algebra, it will be useful to determine
which tangle field theories arising from quantum group representations
admit inner products for which the tangle group action
is unitary, and to calculate $a^\ast$ in these theories.
These reductions of $\H$ are especially interesting
because they have many tantalizing connections with conformal field theory and
3-dimensional topological quantum field theories \cite{Crane,FK}.

As a further generalization, it would be interesting to consider
theories based on framed tangles admitting some sort of
self-intersections.
This is desirable because only states built from self-intersecting links are
not annihilated by the determinant of the metric.
Indeed, whether states built from
non-self-intersecting links are ``physical'' is a matter of dispute
\cite{Smolin,ARS}, and there has been considerable work on finding
solutions to the Hamiltonian constraint built from links with
intersections \cite{BGP1,BGP2,BP,Husain,JS}.  While this topic is still
not well understood, relevant mathematical techniques for
dealing with links admitting self-intersections have recently
been developed by Vassiliev
\cite{Vassiliev} and, in subsequent work, Bar-Natan \cite{Bar-Natan},
 Birman and Lin \cite{BL}.

Finally, it is worth noting that tangle field theories are closely
related to what knot theorists call ``skein modules,'' and that skein
modules suggest generalizations of tangle field theories to $3$-manifolds
with boundary other than $S^2$.  There is a review article on skein modules
by Hoste and Przytycki \cite{HP}.

\vfill
\end{document}